\documentclass[journal=jacsat,manuscript=article]{achemso}

\usepackage[version=3]{mhchem} 

\author{Benjamin A Scott}
\email{bs594@exeter.ac.uk}
\author{Kieran J Cowan}
\affiliation[University of Exeter]
{Department of Physics, University of Exeter, Exeter}
\author{Fraser Burton}
\affiliation[BT Group plc]
{BT Group plc, Ipswich}
\author{Ken E Evans}
\affiliation[University of Exeter]
{Department of Engineering, University of Exeter, Exeter}
\author{Alexander W Powell}
\affiliation[University of Exeter]
{Department of Physics, University of Exeter, Exeter}

\title[An \textsf{achemso} demo]
  {Light-driven Modulation of 1-bit Metamaterial Unit Cells and Arrays}

\abbreviations{UC,NIR,SM,PPS,PLA,EM}

\begin{document}


 
\begin{abstract}

Shape-morphing materials have the potential to realise simple, tunable telecommunication arrays without the need for complex circuitry. However, existing devices either lack individual cell control, or require an integrated power supply, reducing their beamsteering ability and increasing size, weight and power consumption. This paper presents a simple, photothermally activated 1-bit reflectarray unit cell requiring no integrated power sources. The unit cell consists of dimer elements, which are connected or disconnected via a photothermally activated electrical bridge to achieve modulation. The paper discusses the shape memory materials used for switching, and presents a design for a photothermal switching mechanism. The unit cell is validated by waveguide experiments, showing strong agreement with simulations at the working frequency of 3.7GHz. Finally, a 4-cell array is created. Each cell is activated sequentially, with good agreement between simulation and experiment, demonstrating the ability to control each cell exclusively. This study therefore demonstrates that a 1-bit unit cell can be photothermally configured without complicated electronics or an integrated power supply, opening the door to a new class of shape-morphing reflectarray. 

\end{abstract}

\section{Introduction}

The ever-increasing demand on telecommunications networks has created a need for tunable devices that can steer and redirect signals with minimal power consumption. Reflectarray antennas (RA)'s, are flat surfaces that reflect incoming signals to form a desired field pattern. This effect is created by patterning a surface with sub-wavelength unit cells (UC), each with a specified reflected phase at a desired frequency. By designing the response of each UC to form a prescribed phase distribution across the aperture, interference between cells can shape the reflected radiation as required \cite{Nayeri2015Beam-scanningArt}. Conventional RA's are classified into two categories: passive and dynamic devices. Passive RAs are prefabricated devices that are designed to operate in one configuration \cite{Nayeri2015Beam-scanningArt}. Although this meets the requirements for some scenarios, the lack of reconfigurability is often a significant drawback. Dynamic RAs, often referred to as Reconfigurable Intelligent Surfaces (RIS), can manipulate reflected field patterns post-fabrication, unlike their passive counterparts. Standard approaches to achieve reconfigurability include employing PIN diodes \cite{Victor2014ReconfigurableReview}, varactor diodes \cite{Victor2014ReconfigurableReview,Tayebi2015DynamicAntenna,Hum2005RealizingElements} and MEMS \cite{Zhao2019IntegratingMetadevices}. However, these dynamically controlled devices still require integrated driving circuitry and a constant power supply to operate effectively, adding significant cost, losses, and complexity\cite{Uddin2025TheTechniques,Yang2025AdvancedReview}. 

One method to overcome these drawbacks is to employ shape morphing (SM) techniques, where mechanical motion is used to control the electromagnetic (EM) response. This approach allows for many of the benefits of dynamic RAs, specifically the ability to reconfigure their reflected state, without the drawbacks of costly fabrication and powering. This is especially suitable for applications that require reconfiguration, but not fast switching. Examples of previous SM techniques for RF devices include origami and kirigami \cite{Fuchi2015ResonanceFolding,Liu2014ReconfigurableSystem,Carrara2019ARange,Hayes2014Self-FoldingAntennas,Zhang2020Hexagon-TwistStructures,Phon2021MechanicalMetasurface,Jeong2021Four-DimensionalFunctions,Luo2023ReconfigurableMetasurfaces,Sessions2018InvestigationIncidence,Wang2017Origami-BasedChirality}, magnetised soft materials \cite{Pavlick2025ReconfigurableStructure}, rotation of unit cells \cite{Forte2023ChiralTransmittance,Gagnon2013UsingBeam} or entire arrays \cite{Yang2017AElements}. Recent employment of shape-memory materials \cite{Jeong2022FrequencyPrinting,Mazlouman2011ReconfigurableAlloys,Park2023Shape-MorphingOrigami,HussainShah2024RFReview} and electromagnets \cite{Hu2024AElements,Hu2024AConsumption} have also demonstrated a new approach to RF devices. However, these examples all either require a constant power input, lack individual cell control, or require expensive or bulky motors/integrated electromagnets and circuitry to operate. We propose a novel method to dynamically reconfigure a reflectarray using selective near-infrared (NIR) illumination to produce mechanical deformation of a shape memory material via the photothermal effect. This avoids the above drawbacks associated with existing mechanical arrays, allowing for less complex unit cells without incorporated circuitry, and a driving force independent of the array itself that only needs to be active during switching and can be removed when not required. 
Previous work has exploited light-activated shape-memory materials such as prestrained polystyrene (PPS) to create reconfigurable antennas \cite{Hayes2014Self-FoldingAntennas}, for tuning frequency-selective surfaces \cite{Smellie2023Shape-MorphingElements}, and recent work by the authors has proposed the creation of reconfigurable unit cells using this technique. \cite{Scott2025AReflectarrays}.

\begin{figure}
    \centering
    \includegraphics[width=1\linewidth]{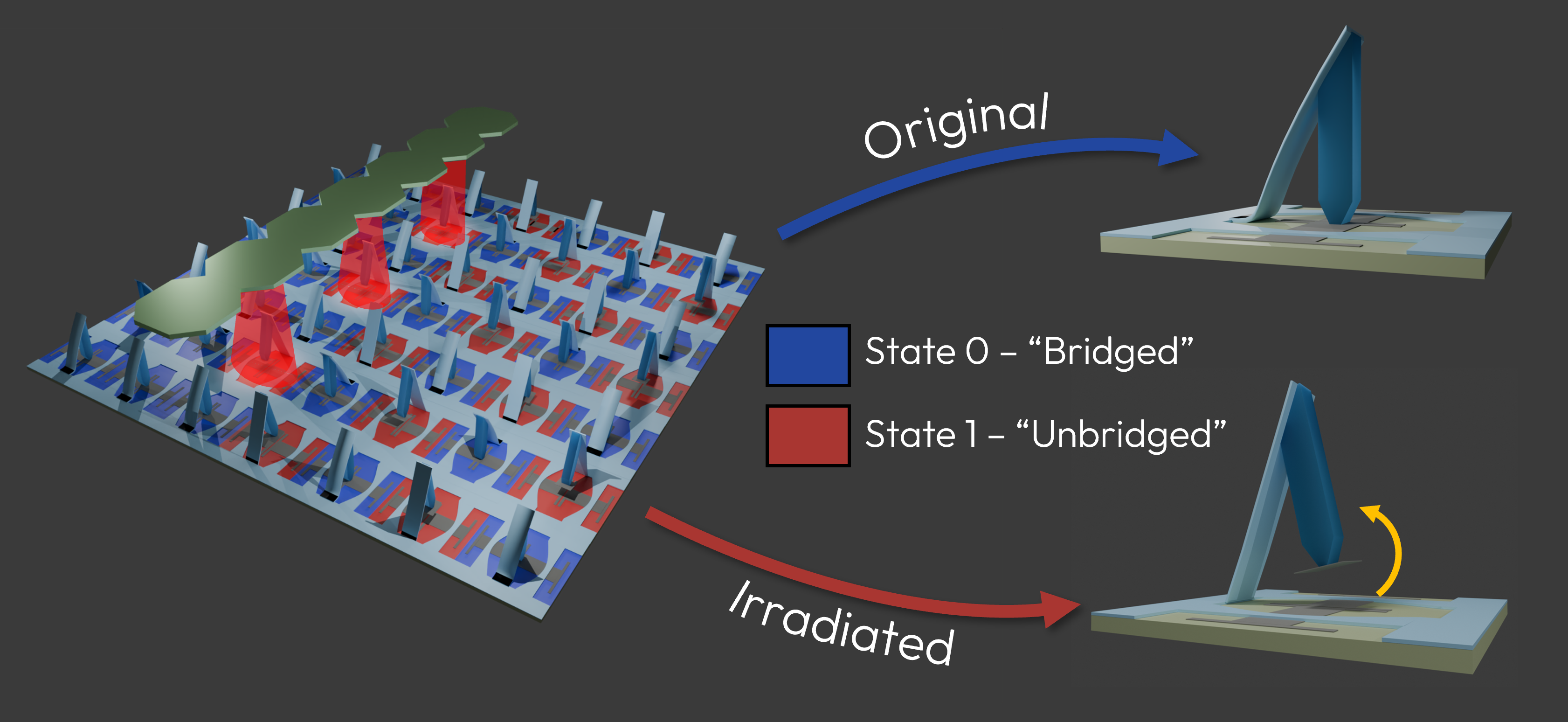}
    \caption{An array of shape-morphing unit cells being activated by an individually controlled array of NIR LEDs.}
    \label{fig:1}
\end{figure}

Herein, we present a fully-optimised microwave RA unit cell activated by near-infrared (NIR) light, capable of 1-bit modulation without embedded electrical biasing or continuous on-cell power within the 5G frequency band at 3.7GHz. The cell employs the shape memory materials PLA and PPS to control an electrical bridge connecting dimer elements (See Fig.\ref{fig:1}). By raising the bridge, the resonance characteristics of the unit cell are altered, resulting in a 1-bit phase modulation between states. This demonstrates a new method of realising dynamic RAs without inbuilt power sources or circuitry, while maintaining the ability for individual cell control. 

\section{Results and discussion}

Creating a shape-morphing UC that can be photothermally activated requires the optimisation of a large parameter space. Firstly, we need to select and refine materials that can be photothermally deformed in a controllable manner. Then we need to decide on an EM structure who's resonant response will be modulated usefully by this deformation. Finally, we must refine the system to overcome practical issues and fabricate it before testing. 

In the first section, the mechanism used to create photothermally activated hinges is discussed; two materials capable of performing this are proposed, and we experimentally demonstrate that selective heating and controlled deformation of these materials. Next, (EM) design of the unit cell is discussed via simulation, including how the 1-bit resonance is switched on and off by employing an electrical bridge between a dimer pair. To achieve this in practice, the electrical bridge must have a force driving it between the dimers. This is optimally achieved by changing the size of the bridging arm and the position of the screws on the sample. Waveguide testing then shows the strong agreement between simulation and experiment. Finally, a 4-cell array is activated using an LED array, with RCS experiments demonstrating the viability of this unit cell in a full array.

\subsection{NIR Activated PPS Hinge}

To achieve cell modulation, we will utilise a shape memory material to create a bilayer hinge, which will deform when illuminated due to photothermal heating. Two shape memory materials are initially considered in this section - prestrained polystyrene (PPS) and polylactic acid (PLA). These materials were chosen because they were commercially available as flat sheets and are known to have strong shape-memory behaviour. Particularly, this paper utilises their thermal expansion coefficient. This is a measure of how much a material expands in response to an increase in temperature. Owing to their shape memory qualities, both materials will begin substantially deforming once the polymer chains begin to relax. The temperature at which the material starts to do this is the glass transition temperature (Tg) and is $\sim$103$^\circ$C and $\sim$60$^\circ$C for PPS and PLA, respectively \cite{Zhang2017OrigamiLight,Liu2017SequentialSheets}. These two materials exert a one-time shape-memory effect, and are therefore not practically reversible. Research is ongoing into reversible shape-memory polymers however, and future work will seek to integrate these into fully reversible designs \cite{Wang2019MultipleApplications}. 

\begin{figure}
    \centering
    \includegraphics[width=0.95\linewidth]{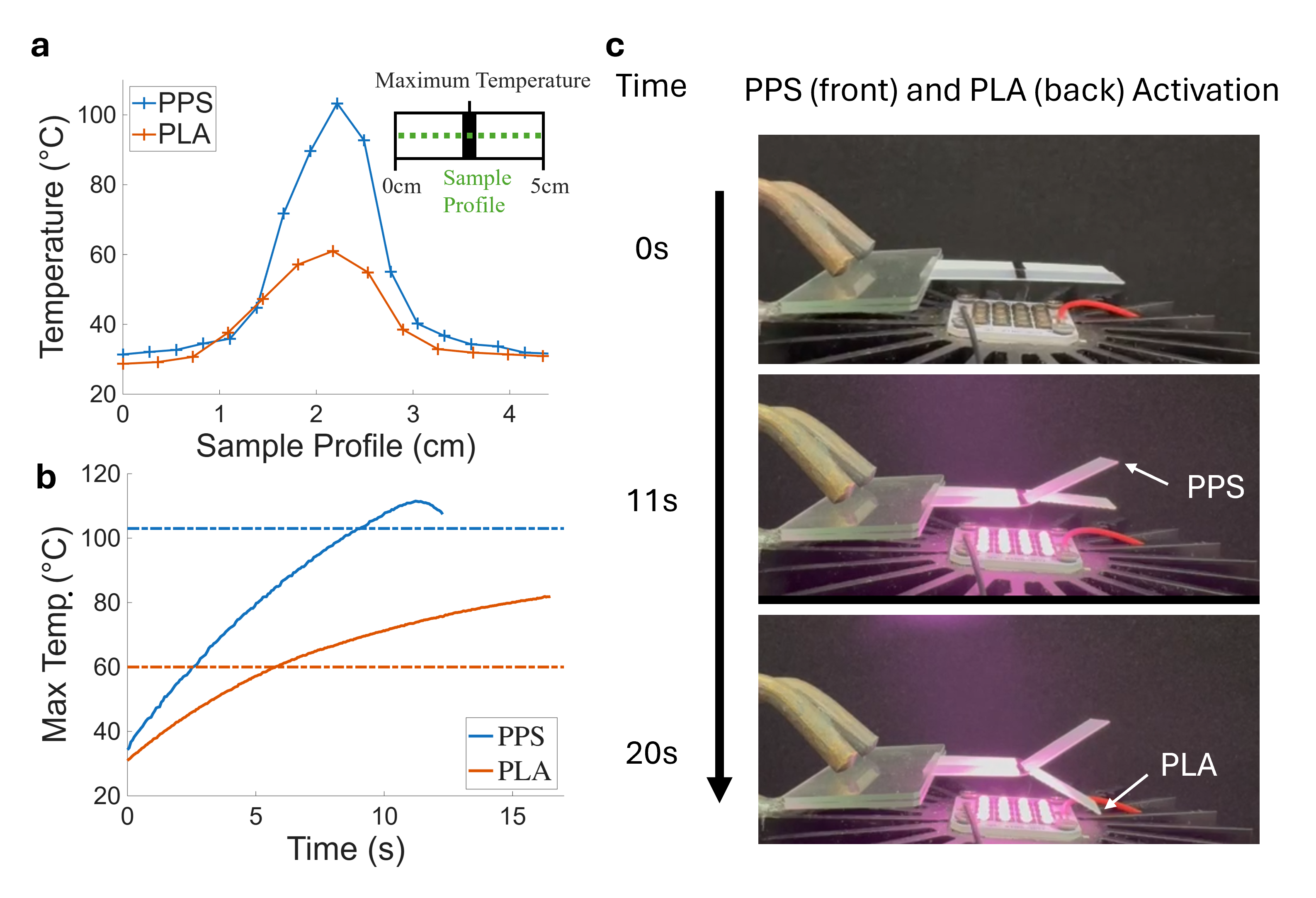}
    \caption{(a) Temperature profile along the long axis of the PPS and PLA sample hinges at Tg temperature as compared to a sample under no illumination and thus at room temperature. (b) The maximum temperature of the PPS and PLA hinges against time with the respective Tg in dashed lines. (c) Photothermal activation of the PPS and PLA hinges in different directions due to the positive and negative thermal expansion coefficients. The absorbing ink is on the upwards facing side of both.}
    \label{fig:2}
\end{figure}

Photothermal heating occurs when absorbed photons are converted into thermal energy. By controlling the absorption rate of the material, localised heating can be achieved. To manipulate this rate of absorption, the authors used black ink to strongly absorb photons emitted from an 850nm OSLON Black PowerCluster LED source. This ink has significantly greater NIR absorption than any of the surrounding PPS material (which does not strongly absorb NIR light), allowing for precise control of localised heating. This is shown in Fig.\ref{fig:2}a, where a temperature profile of a hinge with a central section of black ink is measured at Tg using an Oprtis Xi 400 infrared camera. As shown, there is a far higher temperature generated at the portion of the hinge which is coated in black ink as compared to the rest of the hinge. This trend was the same in multiple experiments, with a representative sample result shown here. An experiment was set up where each material underwent illumination 15 mm above the LED array, with the infrared camera placed above the sample to measure its maximum temperature and an optical camera viewed from the side to see when the material begins to deform. When the footage from these sources was synced, the time taken from illumination to reaching the Tg and finally activation can be found. It was therefore shown that the inked PLA would reach Tg after 5.5s when illuminated 15mm above the LED array, as shown in \ref{fig:2}b from a single representative result. Comparing this to the PPS, which takes 9s to reach Tg, suggests that the PLA may result in a quicker array activation time. In practice, however, the PLA takes much longer to deform once Tg is reached.

As both polymers are poor thermal conductors, and since they are inked on only one side, when they are illuminated, only one side will surpass the glass transition temperature within the measured time. This will create expansion/contraction on only the inked face, generating a bilayer hinging effect. Depending on whether the material expands or contracts above Tg, this will either pull the top faces (the inked side) towards each other or push them away, depending on the values for the rate of expansion on both layers \cite{Zhang2017OrigamiLight,Cui2018ControlledSheet}. For PPS, after reaching Tg, the material undergoes rapid shrinking \cite{Smellie2023Shape-MorphingElements}, while the PLA rapidly expands after reaching its Tg. These characteristics are expected given the prestrained properties of PPS and the typical polymer expansion effect during heating for PLA. Since the PPS shrinks under heating and has a negative thermal expansion coefficient, a photothermally activated PPS bilayer hinge will contract on the face with the black ink and rotate the top faces towards each other. This is compared to a PLA hinge that expands under heating, resulting in the top faces rotating away from each other. This is seen in Fig.\ref{fig:2}c, where both hinges are illuminated partially above this NIR LED source. Supplementary video 1 shows both material hinges under an offset NIR LED array, illustrating this effect. Since both inked faces are facing away from the source this time, and due to the PPS having greater translucence, the PPS reaches Tg first and activates upwards as less light reaches the inked mark for the PLA sample. After another 9s, the PLA hinge activates, but does so in the opposite direction. This shows that controlling the state of a unit cell with a NIR LED source is possible by employing bilayer hinges. Although PLA reached its Tg more rapidly, PPS was selected for the hinge as it provided the required actuation direction and greater mechanical robustness during large deformation. This is further discussed in the following section.

\subsection{Electromagnetic Simulations}

\begin{figure}
    \centering
    \includegraphics[width=1\linewidth]{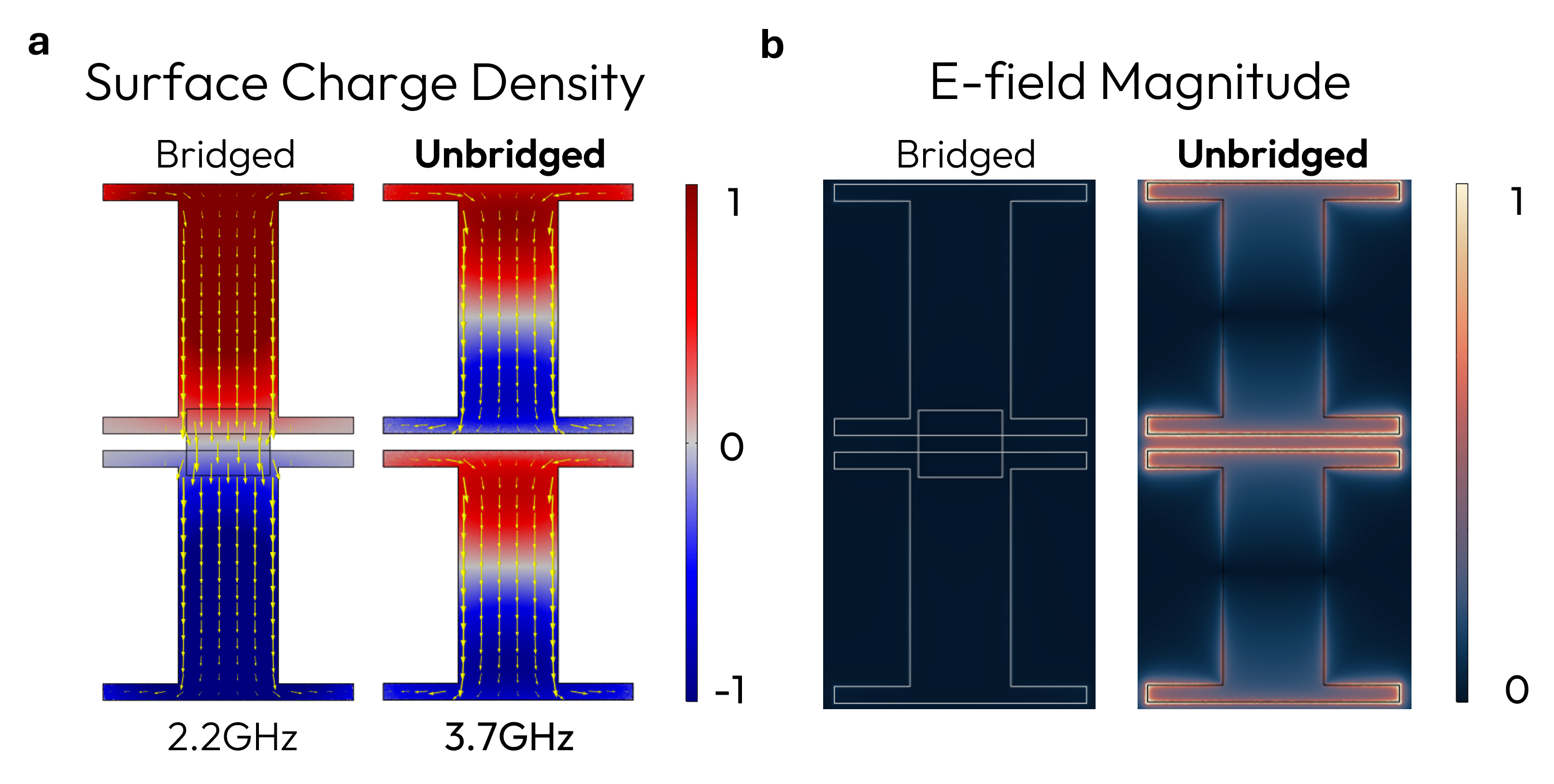}
    \caption{(a) Normalised surface charge for the fundamental modes of the bridged and unbridged elements with proportional surface current density arrows. (b) Normalised electric field of both cell states for the target frequency of 3.7 GHz.}
    \label{fig:4}
\end{figure}

To convert mechanical deformation to EM  modulation,  we will use the shape-morphing hinges described above to switch between electrical states, connecting or disconnecting two adjacent resonant elements to produce a 1-bit system. This design is shown in Fig.\ref{fig:4} and Fig.\ref{fig:5}. UC modulation is made possible due to the large difference in resonant frequency of the fundamental modes for the bridged and unbridged structures. These two different states have a large gap between their respective fundamental resonant frequencies of 2.2GHz and 3.7GHz so there is no interaction between states, and the excitation of the bridged state at the target frequency of 3.7 GHz is negligible. Simulations show a resonance modulation at the desired frequency of 3.7 GHz, which is overlaid with the experimental result in a later section, specifically Fig.\ref{fig:7}c,d. 

Fig.\ref{fig:4}a shows the normalised surface charge density and the normalised surface current of the two cell states at their fundamental resonant frequency. As shown, the bridged elements are attached together to become a single large dipolar element. This is contrasted with the unbridged version, which creates a dimer system. This accounts for the large difference in resonant frequency. 

Simulations were undertaken using COMSOL Multiphysics RF Module, with the substrate relative permittivity modelled as $\varepsilon_r=3.61-0.019i$, which is within the expected error of the published values at 3 GHz. These mimic a waveguide setup, with PEC wall boundary conditions and a TE10 rectangular port excitation. 

When switching between bridged and unbridged states, the electrical bridge should be lifted off the surface of the PCB via the photothermal hinge. To completely decouple the bridge and the element, we found that the hinge should be actuated to an angle greater than 10$^\circ$, where the capacitive effects of the bridge are negligible as verified by FEA.

\subsection{Mechanical Force}

Previously, we have shown that PPS and PLA can be used to move a shape morphing arm under NIR illumination, and that a metallic dimer that can be connected and disconnected can make a 1-bit unit cell for an array. However, to create a stable electrical connection between the dimer elements, sufficient contact pressure is required between the metal bridge and the dimer elements. Without this pressure, the connection is poor, and furthermore, errors and offsets in the manufacture of the hinges and the acrylic posts can create a small angle in the bridge, destroying contact with one of the elements. The required pressure was tested using a digital scale and a multimeter, as shown in Supplementary material 1. The scale was calibrated to 0g with the sample placed on top and multimeter probes on each element. The electrical bridge was then pressed onto the unit cell and, once the multimeter could read a negligible resistance between the probes, showing the bridge had connected the elements, the total weight was read off and converted to pressure. This was tested using 4 electrical bridges over a total of 12 tests. The force required to connect the pair of dimers using the electrical bridge was found on average to be $10.4\mathrm{kPA} \pm 2.5\mathrm{kPa}$.

\begin{figure}
    \centering
    \includegraphics[width=1\linewidth]{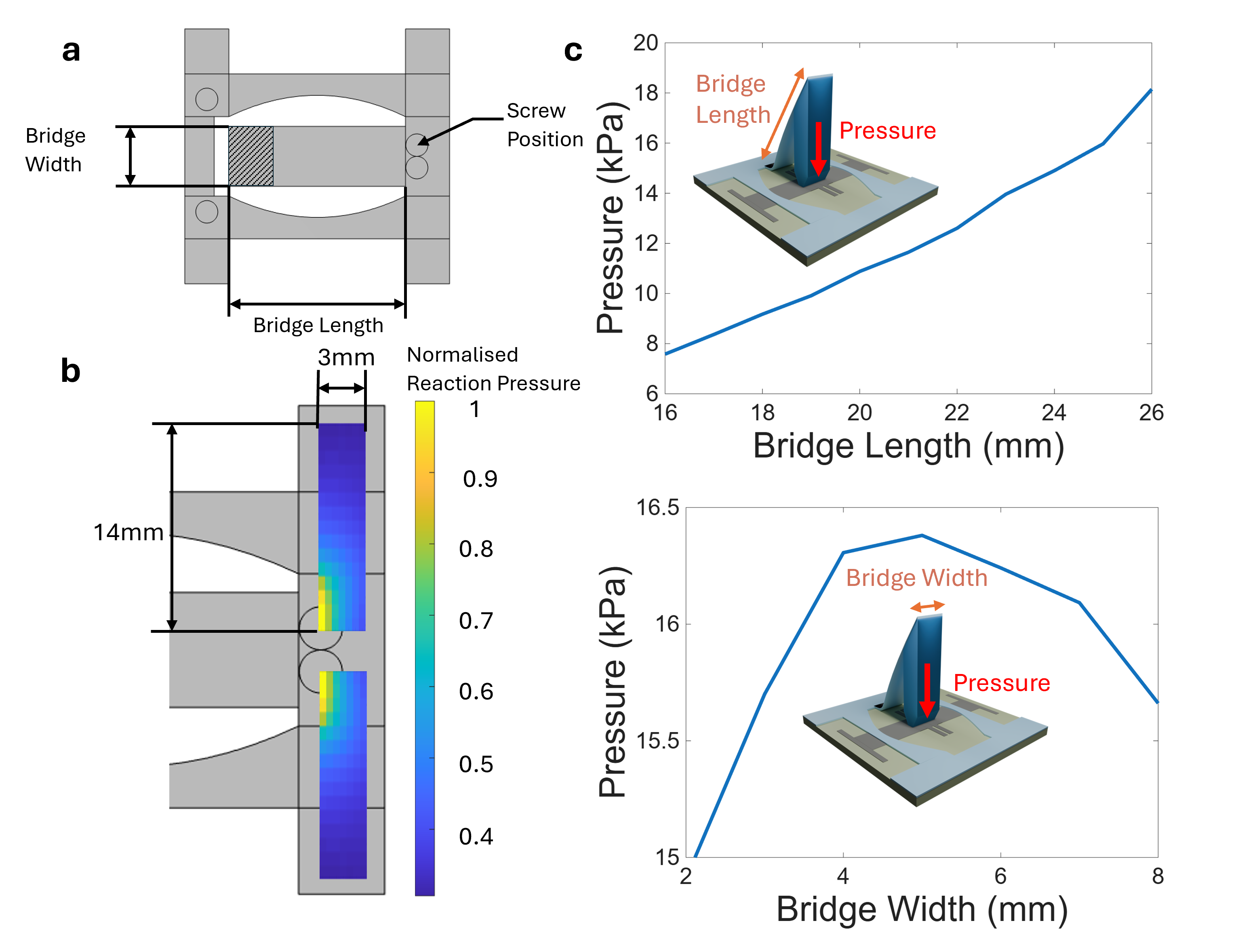}
    \caption{(a) Parameters that affect the reaction pressure of the hinge onto the acrylic post, with the post contact area outlined by a hatched box. (b) Normalised reaction pressure from the hinge as the position of the screw changes. (c) Reaction pressure from the hinge as bridge width and length change for a constant bridge length and width of 24mm and 4mm respectively.}    
    \label{fig:3}
\end{figure}

To produce this pressure, a simple hinge is not sufficient, and so we add an acrylic post to lift the bridging arm above the sample as seen in Fig.\ref{fig:3}c. The reaction force from the lifted hinge bridging arm pushes the acrylic post onto the sample. Therefore, the force pushing down on the electrical bridge is dependent on the width of the bridging arm (with the acrylic post of the same width), the length of the hinging arm (resulting in a longer acrylic post to centre the electrical bridge between the elements), the screw placement on the hinge and finally the material properties of the hinge as described in Fig.\ref{fig:3}a. The hinge material must have shape memory properties, have a high enough Young's modulus to produce the required reaction under deformation, and be flexible enough to undergo large deformation and not relax or break due to the acrylic post, which is examined shortly. As discussed earlier, PPS is used in this paper due to its better photothermal deformation qualities, while PLA was not practical for use in this initial large deformation hinge - the large initial displacement of the material due to acrylic post bending it away from the elements. Mechanical testing shows that it does not have the flexibility to create a working hinge, often breaking or tearing during manipulation. While both have the required shape memory qualities, the PPS has a greater Young's modulus of 897 $\mathrm{MPa}\pm 125\mathrm{MPa}$, as compared to 689 $\mathrm{MPa}\pm 14\mathrm{MPa}$ for PLA. Despite this, the PLA was more brittle and had a tendency to snap when undergoing the deformation required and was thus discounted from further investigation.

To optimise the design to ensure that there is always enough contact pressure as discussed earlier, there are a number of variables to explore using the COMSOL Multiphysics Structural Mechanics Module. Firstly, the placement of the screws closest to the internal arm - these act as a clamping condition to focus the force onto the acrylic post and not dissipate as strain energy through the rest of the hinge. Therefore, the placement of these screws is important to the hinge design. Fig.\ref{fig:3}b shows the normalised pressure of the hinge acting on the acrylic post at different screw distances from the hinging arm. By placing the screws close together and at the base of the internal arm, the reaction pressure is maximised. This pressure quickly decreases as the screws are moved further from the hinging arm, with the pressure halved after $\sim$5mm, making it the greatest factor in the design of reaction force. EM simulations show that the addition of M1.6 PEEK dielectric screws used in this unit cell have a negligible impact on the EM response of the cell. 

Next, the width of the acrylic post and the length of the hinging arm make a difference to the final pressure, as shown Fig.\ref{fig:3}c. An increase in the length of the hinging arm (causing an increase in height of the acrylic post to keep the bridge centred between the elements) greatly increases the reaction pressure from 7.6kPa to 18.2kPa for a length of 16mm to 26mm respectively for a width of 4mm. This trend is quasi-linear as, even with an increasing arm length, the length of the connection to the acrylic arm (as seen in the hatched square in Fig.\ref{fig:2}) does not increase, therefore dividing the force by a constant area. Theoretically, this pressure can be increased further than this in the length of the hinging arm, but are constrained by the system geometry. The length of the hinging arm is limited by the bending of the arm itself and the physical size of the hinge. As the height of the acrylic post is increased to account for the increase in arm length, the angle of the far arm edge begins to rotate towards the post and reduces the force normal to the electrical bridge. This is in contrast to changing the width of the hinging arm, which does change the area of the attached acrylic post as it is increased. As can be seen, for a fixed bridging arm length of 24mm, the pressure starts at 15kPa for a width of 2mm and increases before peaking at 16.4kPa between 4mm and 6mm, before decreasing again. This 4mm to 6mm width range is therefore the most optimal size to maximise applied pressure to the acrylic post and thus the electrical bridge. Therefore, an optimal bridge size of 24mm x 6mm was chosen, corresponding to an acrylic post of size 6mm x 21mm to keep the electrical bridge centred between the elements. The acrylic post was chamfered at the top to better conform to the hinging arm, and bevelled at the bottom, connected to the electrical bridge, to exert greater pressure through the bridge.

\section{Sample Fabrication Method}

\begin{figure}
    \centering
    \includegraphics[width=1\linewidth]{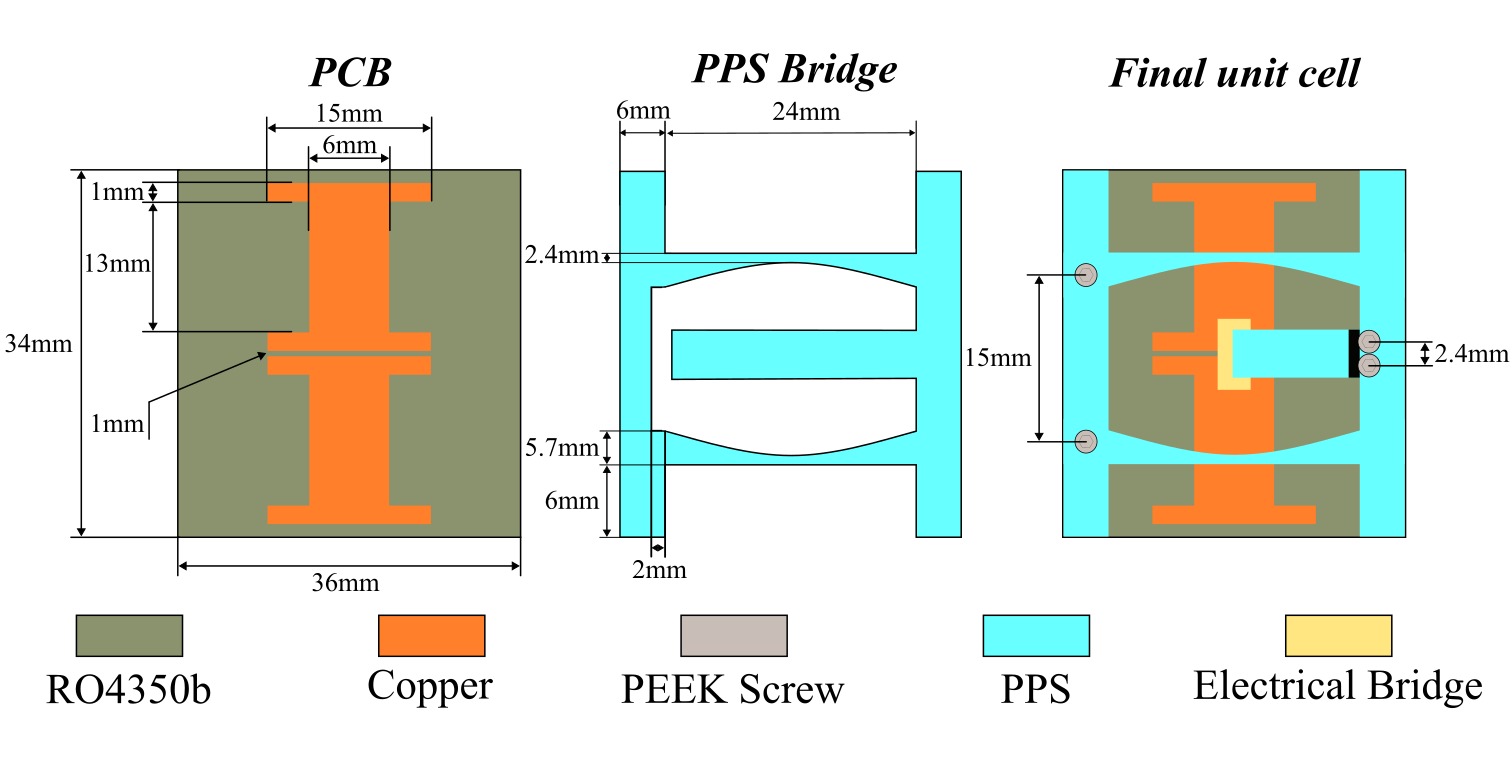}
    \caption{Unit cell schematics of the PCB, PPS bridge and the final UC where both are combined.}
    \label{fig:5}
\end{figure}

The design of the unit cell consists of two dipoles capped with capacitive plates etched onto a 1.524mm Rogers RO4350b substrate. The cell dimensions are 0.42$\lambda$ x 0.44$\lambda$ for the working frequency of 3.7GHz, with cell schematics shown in Fig.\ref{fig:5}. The PPS hinge is designed to cross over the centre of each resonator making up the dimers so as not to interact with the high E-fields at the ends and obscure the tuning of cells' RF characteristics. The design of the resonant elements was chosen for two reasons - to reduce the resonant frequency to one that is in the required 5G frequency band, and to allow for an easy switching mechanism between the two resonant states. This was tuned by altering the size of the capacitive arms and the gap between the elements.

\begin{figure}
    \centering
    \includegraphics[width=1\linewidth]{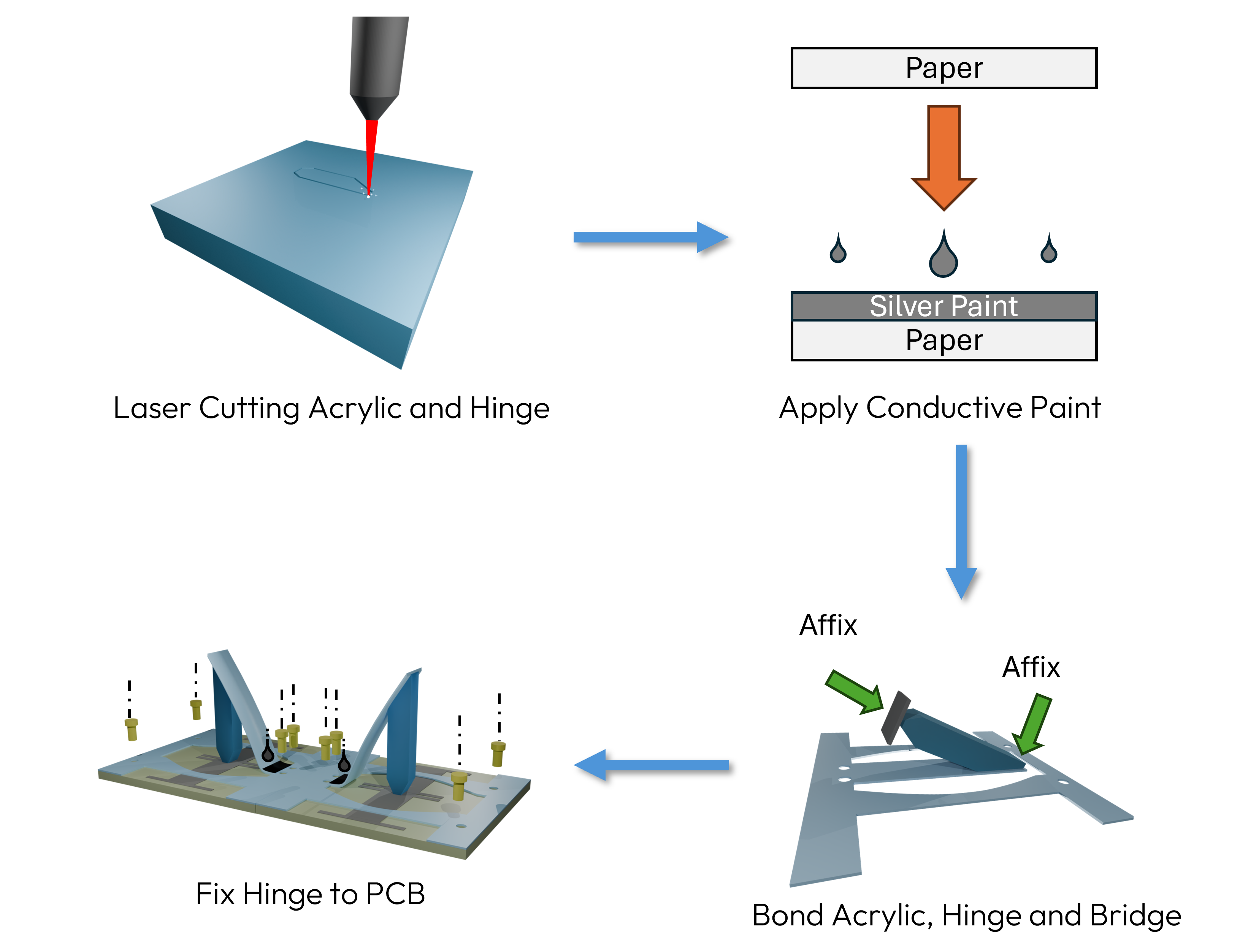}
    \caption{Fabrication process of the mechanical hinge to bridged and unbridged the cell starting with laser cutting the acrylic post and the PPS hinge. Conductive silver paint is applied to printer paper to make an electrical bridge. The electrical bridge is then glued to the bottom of the acrylic post, which itself is glued onto the bridging arm of the hinge. This final hinge is then screwed into the PCB, and black ink is applied to the base of the hinge.}
    \label{fig:6}
\end{figure}

The unit cell PCBs are pre-etched and drilled. Fabrication steps of the mechanical hinge are illustrated in Fig.\ref{fig:6}. To fabricate the hinge for the cell, the PPS base is cut using a 40W Lasertech Pro Range CO2 Laser Cutter at a speed and power of 20mm/s and 10\%, respectively, with a 1cm thick line of black ink black ink at the bottom of the hinging arm, for thermal sensitisation. The laser cutter is also used to cut out the 7mm thick acrylic posts using a speed and power 4.5mm/s and 70\%. To create the electrical bridge, 10mmx7mm of printer paper, is coated with Agar Fast Drying Silver Suspension paint, forming a flexible, highly conductive layer that conforms to the contact surface under pressure and reduces angular errors from the acrylic post. As discussed, a small amount of force is required to create a stable connection between the bridge and the two elements, which arises from the bending of the hinge against the post. The post itself is bonded to the PPS hinge using cyanoacrylate superglue, which itself is screwed into the substrate using M1.6 PEEK screws. To activate the cells, a NIR 16.48W OSLON Black PowerCluster (850nm) LED array is used.

\subsection{Experimental Results}

\begin{figure}
    \centering
    \includegraphics[width=1\linewidth]{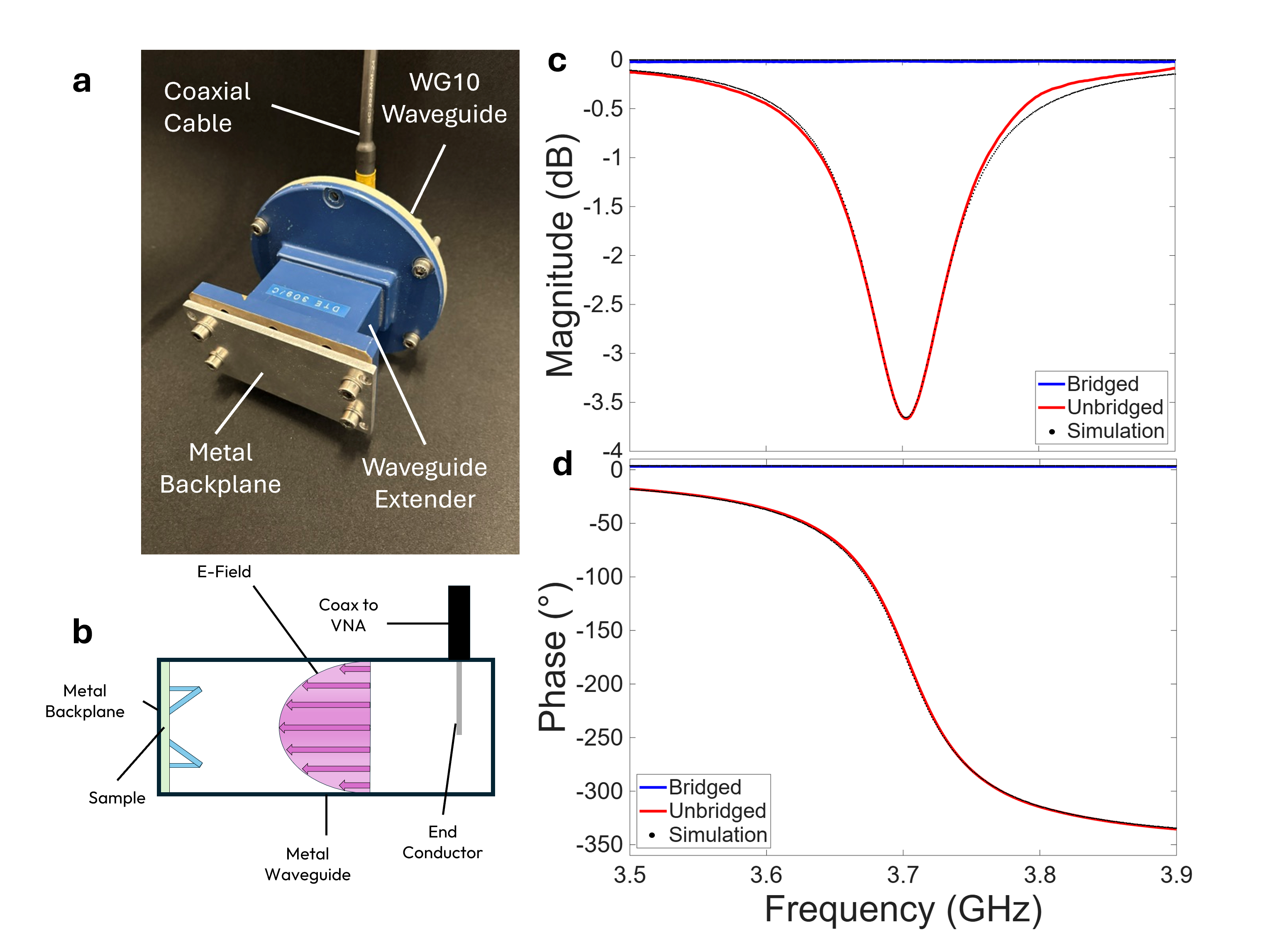}
    \caption{(a) Setup for the waveguide simulator. (b) Schematic of the waveguide. (c)  S11 Magnitude and phase for experiment and simulations.}
    \label{fig:7}
\end{figure}

The simulation results are confirmed experimentally through the employment of the waveguide method. This method allows for experimentally approximating periodic structures using few unit cells, as the electrical field is constrained by the PEC walls as depicted in Fig.\ref{fig:7}b. The measurements were taken in a WG10 waveguide simulator and an Anritsu S820E Sitemaster VNA, as illustrated in Fig.\ref{fig:7}a. Initially, the unit cells in the sample were in the bridged configuration and were measured in the waveguide simulator. Once the result was obtained, the sample was removed, and both cells were activated using NIR LEDs as discussed above. The now unbridged sample was then reinserted into the waveguide simulator for testing.

Using this method, there is strong agreement between experiment and simulation for the cell modulation, as shown in Fig.\ref{fig:7}c. There is a phase flip of ~320$^\circ$ between 3.5GHz and 3.9GHz, with an expected dip in magnitude of 3.65dB due to concentration of electric field energy within the dielectric substrate and associated dielectric losses. Therefore, by working at the target frequency of 3.7GHz, a 1-bit modulation of the cell can be achieved.

Following from this, the next step is to demonstrate that we can create an array of unit cells that can be turned on selectively using a light source. To accomplish this, the NIR LED array was attached to the arm of a robotic stage, in front of the 4x1 cell array, allowing the LED source to be positioned over the cells. For this linear array, the cells were turned on sequentially, one after the other, by moving the LED array above each element and illuminating it for 50s, as shown in Supplementary video 2. The cells can be activated individually by the light source, and the order of unit cells activated can differ depending on the required setup, demonstrating the capability for full 1-bit array programming. 

\begin{figure}
    \centering
    \includegraphics[width=1 \linewidth]{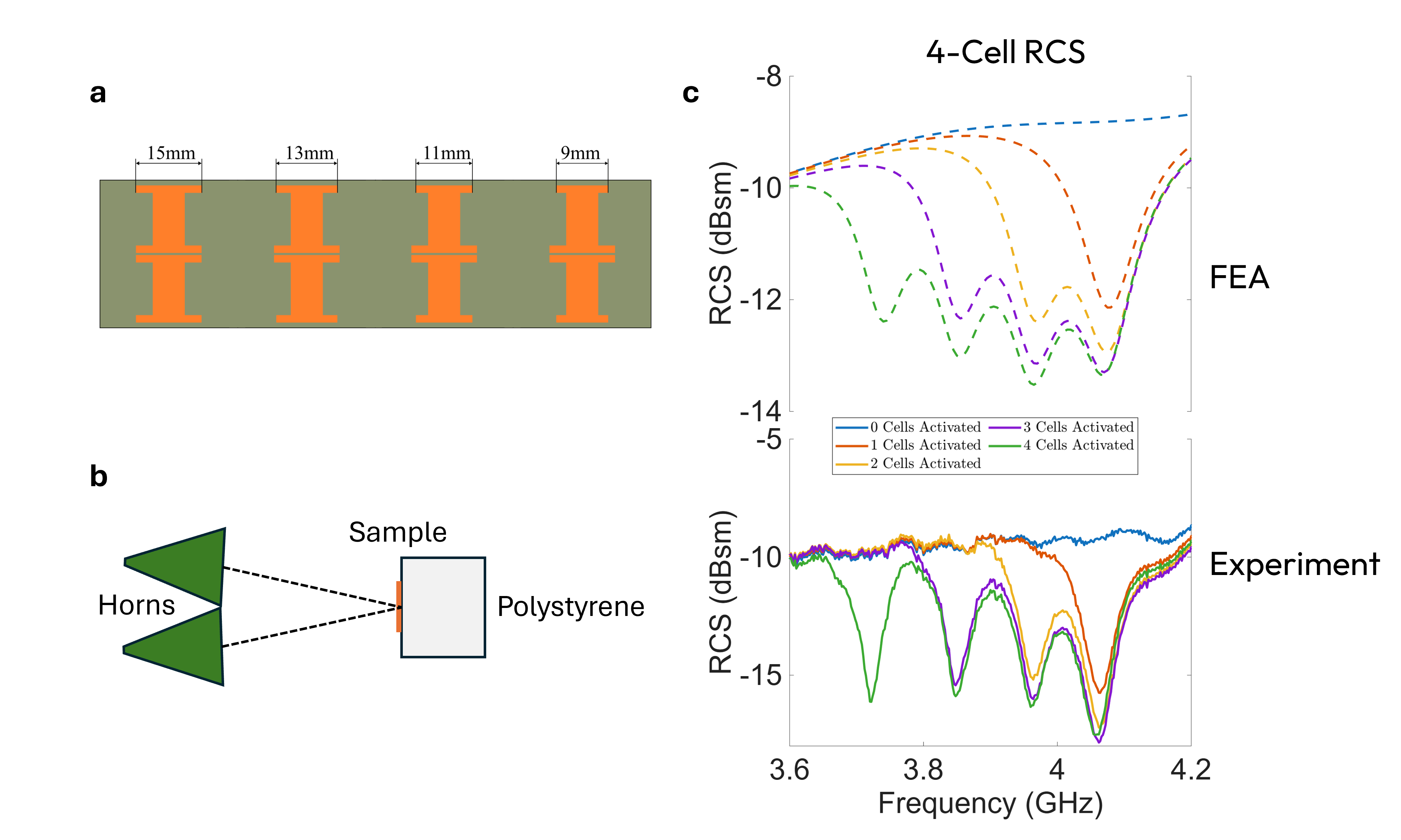}
    \caption{(a) Changing capacitive arm lengths for the 4x1 array. (b) Experimental quasimonostatic RCS measurement setup. (c) RCS experimental and simulation results for different photothermally activated cell configurations. The cells are activated individually starting with the 15mm capacitive plateand getting increasingly smaller.}
    \label{fig:8}
\end{figure}

To demonstrate the possibilities of this in terms of EMs, a quasimonostatic RCS measurement was taken of a 4-cell array at different cell configurations, as shown in Fig.\ref{fig:8}a. By tuning the length of the capacitive arms of the dimers, the resonant frequency can be shifted within the 5G band. This allows the EM measurement of the activation of each individual cell. The measurements were taken in an anechoic chamber using two Flann Microwave dual-polarised horns (Model DP240-AB) and an Anritsu Shockline MS46122B 20 GHz Vector Network Analyzer with the setup shown in Fig.\ref{fig:8}b. The cells were all set up unactivated while an RCS measurement was obtained. This was followed by an activation of a selected cells one-by-one, via the process described above, which was then measured. The results are shown in Fig.\ref{fig:8}c, with each cell in the 4-cell array turned on sequentially. This is compared with COMSOL RF simulations, which show a similar trend between simulation and experiment. As is seen, the RCS measurement of the unit cell being turned on creates a distinct response as compared to the other array configurations, thus showing the ability for automated individual cell control in a system like that shown in Fig.\ref{fig:1}.

\section{Conclusion}

This work demonstrates a pathway toward light-controlled reconfigurable metasurfaces without embedded electrical biasing or continuous on-cell power. A 1-bit modulating RA unit cell is actuated via an NIR LED array for both a single unit cell and a 4x1 array, without the need for integrated circuitry or a constant draw from a power source. The unit cell consists of two dipoles capped with capacitive plates with an electrical bridge in between. Two examples of shape-memory materials are characterised and shown to achieve mechanical hinging of the electrical bridge under NIR illumination. Waveguide experiments show strong agreement with COMSOL simulations, with a clear controllable phase flip at the target frequency 3.7GHz and a loss of less than 4 dB. Following this, a 4x1 array of elements was fabricated and activated sequentially using a robotic stage. Quasimonostatic RCS experiments show that this process can selectively activate individual cells with a clear scalability towards larger arrays. This study has therefore proven that a unit cell can be tuned without complicated electronics or an integrated power supply, opening the door for future work to incorporate further shape morphing techniques in the area of microwave telecommunications. Future work should focus on faster switching, reversible or resettable actuation, and extension to two-dimensional arrays with measured radiation-pattern control.

\begin{acknowledgement}

BA Scott acknowledges financial support from the Engineering and Physical Sciences Research Council (EPSRC) of the United Kingdom and British Telecom PLC, via the EPSRC ICASE Grant No. EP/X524906/1. KJ Cowan acknowledges financial support by an ICASE from the Engineering and Physical Sciences Re-
search Council (EPSRC) and QinetiQ (Grant num-
ber EP/X524906/1, voucher number 220181). AW Powell  acknowledges financial support from a Royal Academy of Engineering research fellowship (RF/201920/19/256).

For the purpose of open access, the author has applied a creative Commons Attribution (CC BY) licence to any Author Accepted Manuscript version arising from this submission.

\end{acknowledgement}



\bibliography{references}

@article{Scott2025AReflectarrays,
    title = {{A 1-Bit Shape-Morphing Unit Cell Design for Photothermally Reconfigurable Reflectarrays}},
    year = {2025},
    journal = {2025 International Conference on Electromagnetics in Advanced Applications (ICEAA)},
    author = {Scott, Benjamin Alistair and Burton, Fl and Evans, K E and Powell, Alex W},
    pages = {535--538},
    url = {https://api.semanticscholar.org/CorpusID:284358659}
}

@article{Yang2017AElements,
    title = {{A Broadband High-Efficiency Reconfigurable Reflectarray Antenna Using Mechanically Rotational Elements}},
    year = {2017},
    journal = {IEEE Transactions on Antennas and Propagation},
    author = {Yang, Xue and Xu, Shenheng and Yang, Fan and Li, Maokun and Hou, Yangqing and Jiang, Shuidong and Liu, Lei},
    number = {8},
    month = {8},
    pages = {3959--3966},
    volume = {65},
    publisher = {Institute of Electrical and Electronics Engineers Inc.},
    doi = {10.1109/TAP.2017.2708079},
    issn = {0018926X}
}

@inproceedings{Carrara2019ARange,
    title = {{A Deployable and Reconfigurable Origami Antenna for Extended Mobile Range}},
    year = {2019},
    booktitle = {2019 IEEE International Symposium on Antennas and Propagation and USNC-URSI Radio Science Meeting},
    author = {Carrara, G P and Russo, N E and Zekios, C L and Georgakopoulos, S V},
    pages = {453--454},
    isbn = {1947-1491},
    doi = {10.1109/APUSNCURSINRSM.2019.8889294}
}

@article{Hu2024AElements,
    title = {{A Highly Efficient 1-bit Reflectarray Antenna Using Electromagnet-Controlled Elements}},
    year = {2024},
    journal = {IEEE Transactions on Antennas and Propagation},
    author = {Hu, A and Konno, K and Chen, Q and Takahashi, T},
    number = {1},
    pages = {506--517},
    volume = {72},
    doi = {10.1109/TAP.2023.3324457},
    issn = {1558-2221}
}

@article{Hu2024AConsumption,
    title = {{A Novel Electromagnet-Controlled Reflectarray Element Employing 1-Bit Height Locking System with Low Power Consumption}},
    year = {2024},
    journal = {IEICE Communications Express},
    author = {Hu, A and Konno, K and Chen, Q},
    number = {7},
    pages = {285--288},
    volume = {13},
    doi = {10.23919/comex.2024XBL0044},
    issn = {2187-0136}
}

@article{Yang2025AdvancedReview,
    title = {{Advanced Metasurface-Based Antennas: A Review}},
    year = {2025},
    journal = {IEEE Open Journal of Antennas and Propagation},
    author = {Yang, W and Li, J and Chen, D and Cao, Y and Xue, Q and Che, W},
    number = {1},
    pages = {6--24},
    volume = {6},
    doi = {10.1109/OJAP.2024.3465513},
    issn = {2637-6431}
}

@article{Nayeri2015Beam-scanningArt,
    title = {{Beam-scanning reflectarray antennas: A technical overview and state of the art}},
    year = {2015},
    journal = {IEEE Antennas and Propagation Magazine},
    author = {Nayeri, Payam and Yang, Fan and Elsherbeni, Atef Z.},
    number = {4},
    month = {8},
    pages = {32--47},
    volume = {57},
    publisher = {IEEE Computer Society},
    doi = {10.1109/MAP.2015.2453883},
    issn = {10459243}
}

@article{Forte2023ChiralTransmittance,
    title = {{Chiral Mechanical Metamaterials for Tunable Optical Transmittance}},
    year = {2023},
    journal = {Advanced Functional Materials},
    author = {Forte, Antonio Elia and Melancon, David and Zanati, Mohamed and De Giorgi, Marta and Bertoldi, Katia},
    publisher = {John Wiley and Sons Inc},
    doi = {10.1002/adfm.202214897},
    issn = {16163028}
}

@article{Cui2018ControlledSheet,
    title = {{Controlled bending and folding of a bilayer structure consisting of a thin stiff film and a heat shrinkable polymer sheet}},
    year = {2018},
    journal = {Smart Materials and Structures},
    author = {Cui, Jianxun and Adams, John G M and Zhu, Yong},
    number = {5},
    volume = {27},
    url = {https://doi.org/10.1088/1361-665X/aab9d9},
    doi = {10.1088/1361-665X/aab9d9}
}

@article{Tayebi2015DynamicAntenna,
    title = {{Dynamic Beam Shaping Using a Dual-Band Electronically Tunable Reflectarray Antenna}},
    year = {2015},
    journal = {IEEE Transactions on Antennas and Propagation},
    author = {Tayebi, Amin and Tang, Junyan and Roy Paladhi, Pavel and Udpa, Lalita and Udpa, Satish and Rothwell, Edward},
    month = {8},
    volume = {63},
    doi = {10.1109/TAP.2015.2456939}
}

@article{Jeong2021Four-DimensionalFunctions,
    title = {{Four-Dimensional Printed Shape Memory Metasurface to Memorize Absorption and Reflection Functions}},
    year = {2021},
    journal = {ACS Applied Materials {\&} Interfaces},
    author = {Jeong, Heijun and Park, Eiyong and Lim, Sungjoon},
    number = {49},
    month = {12},
    pages = {59487--59496},
    volume = {13},
    publisher = {American Chemical Society},
    url = {https://doi.org/10.1021/acsami.1c17968},
    doi = {10.1021/acsami.1c17968},
    issn = {1944-8244}
}

@article{Jeong2022FrequencyPrinting,
    title = {{Frequency memorizing shape morphing microstrip monopole antenna using hybrid programmable 3-dimensional printing}},
    year = {2022},
    journal = {Additive Manufacturing},
    author = {Jeong, Heijun and Park, Eiyong and Lim, Sungjoon},
    pages = {102988},
    volume = {58},
    url = {https://www.sciencedirect.com/science/article/pii/S2214860422003815},
    doi = {https://doi.org/10.1016/j.addma.2022.102988},
    issn = {2214-8604}
}

@article{Zhang2020Hexagon-TwistStructures,
    title = {{Hexagon-Twist Frequency Reconfigurable Antennas via Multi-Material Printed Thermo-Responsive Origami Structures}},
    year = {2020},
    journal = {Frontiers in Materials},
    author = {Zhang, Ya-Jing and Wang, Li-Chen and Song, Wei-Li and Chen, Mingji and Fang, Daining},
    volume = {Volume 7 - 2020},
    url = {https://www.frontiersin.org/journals/materials/articles/10.3389/fmats.2020.600863},
    doi = {10.3389/fmats.2020.600863},
    issn = {2296-8016}
}

@article{Zhao2019IntegratingMetadevices,
    title = {{Integrating microsystems with metamaterials towards metadevices}},
    year = {2019},
    journal = {Microsystems and Nanoengineering},
    author = {Zhao, Xiaoguang and Duan, Guangwu and Li, Aobo and Chen, Chunxu and Zhang, Xin},
    number = {1},
    month = {12},
    volume = {5},
    publisher = {Nature Publishing Group},
    doi = {10.1038/s41378-018-0042-1},
    issn = {20557434}
}

@article{Sessions2018InvestigationIncidence,
    title = {{Investigation of fold-dependent behavior in an origami-inspired FSS under normal incidence}},
    year = {2018},
    journal = {Progress In Electromagnetics Research M},
    author = {Sessions, Deanna and Fuchi, Kazuko and Pallampati, Sumana and Grayson, David and Seiler, Steven and Bazzan, Giorgio and Reich, Gregory and Buskohl, Philip and Huff, Gregory},
    month = {1},
    pages = {131--139},
    volume = {63},
    doi = {10.2528/PIERM17092504}
}

@article{Phon2021MechanicalMetasurface,
    title = {{Mechanical and Self-Deformable Spatial Modulation Beam Steering and Splitting Metasurface}},
    year = {2021},
    journal = {Advanced Optical Materials},
    author = {Phon, Ratanak and Kim, Yeonju and Park, Eiyong and Jeong, Heijun and Lim, Sungjoon},
    number = {19},
    month = {10},
    pages = {2100821},
    volume = {9},
    publisher = {John Wiley {\&} Sons, Ltd},
    url = {https://doi.org/10.1002/adom.202100821},
    doi = {https://doi.org/10.1002/adom.202100821},
    issn = {2195-1071}
}

@article{Wang2019MultipleApplications,
    title = {{Multiple and two-way reversible shape memory polymers: Design strategies and applications}},
    year = {2019},
    journal = {Progress in Materials Science},
    author = {Wang, Kaojin and Jia, Yong-Guang and Zhao, Chuanzhuang and Zhu, X X},
    pages = {100572},
    volume = {105},
    url = {https://www.sciencedirect.com/science/article/pii/S0079642519300489},
    doi = {https://doi.org/10.1016/j.pmatsci.2019.100572},
    issn = {0079-6425}
}

@article{Zhang2017OrigamiLight,
    title = {{Origami and kirigami inspired self-folding for programming three-dimensional shape shifting of polymer sheets with light}},
    year = {2017},
    journal = {Extreme Mechanics Letters},
    author = {Zhang, Qiuting and Wommer, Jonathon and O'Rourke, Connor and Teitelman, Joseph and Tang, Yichao and Robison, Joshua and Lin, Gaojian and Yin, Jie},
    month = {2},
    pages = {111--120},
    volume = {11},
    publisher = {Elsevier Ltd},
    doi = {10.1016/j.eml.2016.08.004},
    issn = {23524316}
}

@article{Wang2017Origami-BasedChirality,
    title = {{Origami-Based Reconfigurable Metamaterials for Tunable Chirality}},
    year = {2017},
    journal = {Advanced Materials},
    author = {Wang, Zuojia and Jing, Liqiao and Yao, Kan and Yang, Yihao and Zheng, Bin and Soukoulis, Costas M and Chen, Hongsheng and Liu, Yongmin},
    number = {27},
    month = {7},
    pages = {1700412},
    volume = {29},
    publisher = {John Wiley {\&} Sons, Ltd},
    url = {https://doi.org/10.1002/adma.201700412},
    doi = {https://doi.org/10.1002/adma.201700412},
    issn = {0935-9648}
}

@article{Hum2005RealizingElements,
    title = {{Realizing an electronically tunable reflectarray using varactor diode-tuned elements}},
    year = {2005},
    journal = {IEEE Microwave and Wireless Components Letters},
    author = {Hum, Scan V. and Okoniewski, Michael and Davies, Robert J.},
    number = {6},
    month = {6},
    pages = {422--424},
    volume = {15},
    doi = {10.1109/LMWC.2005.850561},
    issn = {15311309}
}

@article{Mazlouman2011ReconfigurableAlloys,
    title = {{Reconfigurable Axial-Mode Helix Antennas Using Shape Memory Alloys}},
    year = {2011},
    journal = {IEEE Transactions on Antennas and Propagation},
    author = {Mazlouman, S Jalali and Mahanfar, A and Menon, C and Vaughan, R G},
    number = {4},
    pages = {1070--1077},
    volume = {59},
    doi = {10.1109/TAP.2011.2109686},
    issn = {1558-2221}
}

@inproceedings{Liu2014ReconfigurableSystem,
    title = {{Reconfigurable helical antenna based on an origami structure for wireless communication system}},
    year = {2014},
    booktitle = {2014 IEEE MTT-S International Microwave Symposium (IMS2014)},
    author = {Liu, Xueli and Yao, Shun and Georgakopoulos, S V and Cook, B S and Tentzeris, M M},
    month = {7},
    pages = {1--4},
    isbn = {0149-645X},
    doi = {10.1109/MWSYM.2014.6848553}
}

@article{Luo2023ReconfigurableMetasurfaces,
    title = {{Reconfigurable High-Efficiency metadevice using Kirigami-Inspired phase gradient metasurfaces}},
    year = {2023},
    journal = {Results in Physics},
    author = {Luo, Huiling and Wang, Yanzhao and Wang, Mingzhao and Wang, Chaohui and Liu, Tong and Ling, Xiaohui and Xu, He-Xiu},
    month = {10},
    pages = {106949},
    volume = {53},
    doi = {10.1016/j.rinp.2023.106949},
    issn = {22113797}
}

@inproceedings{Pavlick2025ReconfigurableStructure,
    title = {{Reconfigurable Linear-to-Circular Polarization Converter Based on a Tunable Shape Morphing Structure}},
    year = {2025},
    booktitle = {2025 IEEE International Symposium on Antennas and Propagation and North American Radio Science Meeting (AP-S/CNC-USNC-URSI)},
    author = {Pavlick, W and Sim, J and Shapero, T and West, D and Kovitz, J and Zhao, R R and Ghalichechian, N},
    pages = {2931--2934},
    isbn = {1947-1491},
    doi = {10.1109/AP-S/CNC-USNC-URSI55537.2025.11266693}
}

@article{Victor2014ReconfigurableReview,
    title = {{Reconfigurable Reflectarrays and Array Lenses for Dynamic Antenna Beam Control: A Review}},
    year = {2014},
    journal = {IEEE TRANSACTIONS ON ANTENNAS AND PROPAGATION},
    author = {Victor, Sean and Perruisseau-Carrier, Julien},
    number = {1},
    pages = {183},
    volume = {62},
    url = {http://ieeexplore.ieee.org.},
    doi = {10.1109/TAP.2013.2287296}
}

@inproceedings{Fuchi2015ResonanceFolding,
    title = {{Resonance tuning of RF devices through origami folding}},
    year = {2015},
    booktitle = {20th International Conference on Composite Materials},
    author = {Fuchi, Kazuko and Buskohl, Philip R and Joo, James J and Reich, Gregory W and Vaia, Richard A},
    month = {6},
    organization = {20th International Conference on Composite Materials},
    address = {Copenhagen}
}

@misc{HussainShah2024RFReview,
    title = {{RF advancements enabled by smart shape memory materials in the microwave Regime: A state-of-the-art review}},
    year = {2024},
    booktitle = {Materials Today Physics},
    author = {Hussain Shah, Syed Imran and Lim, Sungjoon},
    month = {5},
    volume = {44},
    publisher = {Elsevier Ltd},
    doi = {10.1016/j.mtphys.2024.101435},
    issn = {25425293}
}

@article{Hayes2014Self-FoldingAntennas,
    title = {{Self-Folding Origami Microstrip Antennas}},
    year = {2014},
    journal = {IEEE TRANSACTIONS ON ANTENNAS AND PROPAGATION},
    author = {Hayes, Gerard J and Liu, Ying and Genzer, Jan and Lazzi, Gianluca and Dickey, Michael D},
    number = {10},
    volume = {62},
    url = {http://www.ieee.org/publications_standards/publications/rights/index.html},
    doi = {10.1109/TAP.2014.2346188}
}

@article{Liu2017SequentialSheets,
    title = {{Sequential self-folding of polymer sheets}},
    year = {2017},
    journal = {Science Advances},
    author = {Liu, Ying and Shaw, Brandi and Dickey, Michael D. and Genzer, Jan},
    number = {3},
    month = {3},
    volume = {3},
    url = {https://www.science.org/doi/10.1126/sciadv.1602417},
    doi = {10.1126/sciadv.1602417},
    issn = {2375-2548}
}

@article{Park2023Shape-MorphingOrigami,
    title = {{Shape-Morphing Antenna Array by 4D-Printed Multimaterial Miura Origami}},
    year = {2023},
    journal = {ACS Applied Materials {\&} Interfaces},
    author = {Park, Seyeon and Park, Eiyong and Lee, Minjae and Lim, Sungjoon},
    number = {42},
    month = {10},
    pages = {49843--49853},
    volume = {15},
    publisher = {American Chemical Society},
    url = {https://doi.org/10.1021/acsami.3c11425},
    doi = {10.1021/acsami.3c11425},
    issn = {1944-8244}
}

@inproceedings{Smellie2023Shape-MorphingElements,
    title = {{Shape-Morphing Polymers for Tunable Frequency Selective Surfaces and Reflectarray Elements}},
    year = {2023},
    booktitle = {2023 SBMO/IEEE MTT-S International Microwave and Optoelectronics Conference (IMOC)},
    author = {Smellie, Daanish and Scott, Benjamin Alistair and Powell, Alex W},
    month = {11},
    pages = {97--99},
    publisher = {IEEE},
    isbn = {979-8-3503-2067-1},
    doi = {10.1109/IMOC57131.2023.10379691}
}

@article{Uddin2025TheTechniques,
    title = {{The Evolutions of Reflectarray Antennas for Small Satellites: Navigating to Design, Fabrication, and Performance Techniques}},
    year = {2025},
    journal = {IEEE Access},
    author = {Uddin, Jasim and Nagi, Harvinder Sing and Falkner, Benjamin and Akhavan, Reza and Rajan, Ginu and Platts, Jon and Hewage, Chaminda T E R and Uggalla, Leshan},
    pages = {156465--156485},
    volume = {13},
    doi = {10.1109/ACCESS.2025.3597389}
}

@article{Gagnon2013UsingBeam,
    title = {{Using Rotatable Planar Phase Shifting Surfaces to Steer a High-Gain Beam}},
    year = {2013},
    journal = {IEEE Transactions on Antennas and Propagation},
    author = {Gagnon, Nicolas and Petosa, Aldo},
    number = {6},
    month = {6},
    pages = {3086--3092},
    volume = {61},
    doi = {10.1109/TAP.2013.2253298},
    issn = {0018-926X}
}

\end{document}